\documentclass[rnote,traditabstract]{aa} % for the research notes
\usepackage{color}
\usepackage{graphicx}

\usepackage{txfonts}

\usepackage{natbib}

\begin{document}
  \title{Dense Molecular Gas around AGN: HCN/CO in NGC\,3227\thanks{Based on observations carried out with the IRAM Plateau de Bure Interferometer. IRAM is supported by INSU/CNRS (France), MPG (Germany) and IGN (Spain).}}

   \author{R. Davies\inst{1}
          \and
          D. Mark\inst{2}
          \and
          A. Sternberg\inst{2}
          }

   \institute{Max Planck Institut f\"ur extraterrestrische Physik, Postfach 1312, 85741 Garching, Germany\\
              \email{davies@mpe.mpg.de}
         \and
             Sackler School of Physics and Astronomy, Tel Aviv University, Tel Aviv 69978, Israel}

   \date{Received ...; accepted ...}

% \abstract{}{}{}{}{} 

% 5 {} token are mandatory

% context, aims, methods, results, conclusions

\abstract
{There is now convincing evidence that the intensity of HCN molecular line emission is enhanced around active galactic nuclei. In this paper we examine the specific case of the Seyfert galaxy NGC\,3227, for which there are subarcsecond resolution data for the HCN (1-0) 88 GHz
and CO (2-1) 230 GHz rotational lines, enabling us to spatially separate a
circumnuclear ring at a radius of 140\,pc and an inner nuclear region within 40\,pc of the AGN. The HCN(1-0)/CO(2-1) flux ratio differs by more than an order of magnitude between these two regions. We carry out large velocity gradient (LVG) computations to determine the range of parameters (gas temperature and density, HCN/CO abundance ratio, column densities and velocity gradients) that yield physically plausible solutions for the observed flux ratio in the central 100\,pc.  The observed HCN/CO intensity ratio 
in the nucleus is consistent with very optically thick thermalized emission in very dense ($\gtrsim 10^5$~cm$^{-3}$)
gas, in which case the HCN/CO abundance ratio there
is unconstrained. Alternatively, the HCN/CO intensity ratio could be due to optically thinner emissions but with
very high ($\sim 10^{-2}$) HCN/CO abundance ratios. This possibility is more consistent with the CO and HCN emissions seen in the 
nuclei of the Seyfert galaxies NGC\,1068 and NGC\,6951. 
It would imply the velocity gradients are large and the clouds may be gravitationally unbound.
%We show there is one region of the parameter space that can accommodate a number of issues: the flux ratio is close to the optically thick limit, the clouds are not too far from the self-gravitating regime, and the same set of physical conditions can explain the HCN(1-0)/CO(2-1) flux ratios observed on similar scales in two other galaxies.
%These clouds are gravitationally unbound.
We estimate that the X-ray ionisation rate at radii less than 20 pc in the centre of NGC 3227 exceeds $10^{-13}$~s$^{-1}$.
X-ray ionisation and heating may lead to high HCN/CO ratios in warm gas in a high-ionisation molecular phase near the AGN.
%Finally, we show that the x-ray ionisation rate due to the AGN could be responsible for the high HCN/CO line intensity ratio in the central region.
}


\maketitle
%
%________________________________________________________________

\section{Introduction}
\label{sec:intro}

During the last decade there has been an increasing observational effort to understand the nature of the HCN emission from galaxies.
This includes studies of HCN emission in nearby AGN and starbursts \citep{koh01,koh05,koh08,kri07,kri08} as well as luminous and ultraluminous galaxies \citep{gra06,gra08}. One result is that HCN line emission is enhanced with respect to CO in at least some AGN compared with starbursts \citep{koh05,koh08}.  The high HCN/CO intensity ratios could be due to high molecular gas densities near the AGN, and/or high
HCN/CO abundance ratios that might be due to elevated X-ray ionisation and heating rates near the accreting black holes. 
%Such an effect is expected, at least qualitatively, since theoretical models of X-ray irradiation of gas suggest that the equilibrium abundances are modified -- and specifically that of HCN increased -- by the higher electron abundance resulting from X-ray secondary ionisation \citep{kro83,lep96,mal96,mei05,bog05,mei07}.
%In this respect, the best studied object is NGC\,1068 for which \cite{ste94} showed that HCN had a remarkably high abundance.
For example, in an early study,  \cite{ste94} concluded that the large HCN/CO intensity ratio observed in the nuclear ($< 100$ pc) region
in the Seyfert-2 galaxy NGC 1068 indicates a high HCN/CO abundance ratio $\sim 10^{-2}$ near the AGN. 
\cite{lep96} and \cite{use04} argued that elevated HCN/CO abundance ratios are signatures of high X-ray ionisation rates \citep{mal96,mei05,bog05,mei07}.  High gas densities
also likely play a role in boosting the HCN line intensity \citep{kri08}.

%In this paper, we present the first results of a quantitative study of molecular line ratios at high ($\sim$1\arcsec) spatial resolution.
%Crucially, we are able to spatially resolve line emission in the central tens of parsecs, and distinguish it from the circumnuclear emission on scales of hundreds of parsecs.
%Our goal is to observationally constrain the physical properties of the molecular gas close around AGN; and to understand, with reference to theoretical models of X-ray irradiation and molecular abundance, the variations of line intensity.
In this paper we present high resolution observations of CO (2-1) and HCN (1-0)  rotational line emissions in the inner regions of
NGC\,3227, a nearby (D=17\,Mpc; 1\arcsec $\sim$ 80\,pc) Seyfert galaxy for which detailed studies have been made of the stellar \citep{dav06,dav07} and gaseous \citep{sch00,hic09} content of its central regions.  We analyze the data using
large-velocity-gradient (LVG) computations.
For this galaxy we again find a significant enhancement in the HCN/CO intensity ratio close to the active nucleus.
The observed nuclear intensity ratio is consistent with optically thick thermalized emission in dense ($\gtrsim 10^5$~cm$^{-3}$) gas.
Alternatively, and especially in comparison with similar data in NGC 1068 and NGC 6951, the nuclear emissions could be
tracing optically thinner emission in which the HCN/CO abundance ratio is large.
%In a companion paper, \cite{san11} analyse the kinematics of the HCN in NGC\,3227 together with several other galaxies to provide additional independent constraints on the gas properties.

%______________________________________________________________

\section{Observations}
\label{sec:data}

The analysis in this paper is based on subarcsecond resolution observations of the CO(2-1) 230.5\,GHz and HCN(1-0) 88.6\,GHz lines.
The CO(2-1) data, for which the beam is 0.6\arcsec, were previously presented by \cite{sch00}.
New 3\,mm HCN(1-0) data, with a 0.9\arcsec$\times$1.2\arcsec\ beam, were obtained during February 2009 in the A configuration (760\,m baseline) of the six 15-meter antennas of the IRAM Plateau de Bure Interferometer.
The H$^{12}$CN(1-0) line at 88.6\,GHz and the H$^{13}$CN(1-0) line at 86.3\,GHz were observed together, using a single polarisation for each 1\,GHz bandwidth segment.
The system temperature was 80--100\,K.
Atmospheric conditions were moderate, with winds and $\sim5$\,mm precipitable water vapour.
Phase and amplitude variations were calibrated out by interleaving
reference observations of standard calibration sources.
The data were processed and calibrated using the CLIC program in the
IRAM GILDAS package, and binned spectrally to a resolution of 
50\,km$^{-1}$ to increase the signal to noise.
The kinematics of the HCN(1-0) line emission in NGC\,3227 are analysed by \cite{san11} together with several other galaxies.

\begin{figure}
\centering
\includegraphics[angle=0,width=8.8cm]{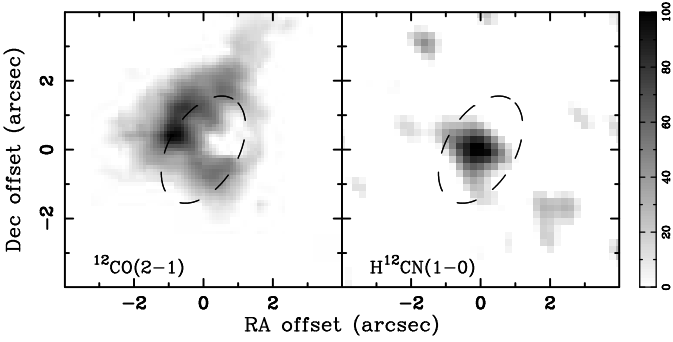}
\caption{\label{fig:obs}
Images of the CO(2-1) line emission \citep[from][]{sch00} and HCN(1-0) line emission from NGC\,3227.
Overdrawn is the ellipse tracing the circumnuclear ring \citep{dav06}.
While the CO emission originates mostly from the ring, the HCN intensity is far higher in the central region. There is an order-of-magnitude change in the flux ratio between these regions.
}
\end{figure}

%One of the remarkable differences between these lines is that, 
It is remarkable that despite the similar beam sizes, the CO(2-1) map clearly shows the circumnuclear ring at a radius of 140\,pc (1.7\arcsec), while the HCN map shows emission from only the central region, within a few tens of parsecs from the nucleus.
This can be seen clearly in Fig.~\ref{fig:obs} and is reflected in the line fluxes given in Table.~\ref{tab:data}.
In the nuclear region, which we define here to be the central 1\arcsec\ (i.e. $r<40$\,pc), the flux ratio is F$_{HCN1-0}$/F$_{CO2-1}$=0.11;
on the other hand in the circumnuclear ring, for which we take an annulus 2--4\arcsec, we find F$_{HCN1-0}$/F$_{CO2-1}$=0.01, an order of magnitude less.
Since the emission is spatially resolved in both cases, this difference cannot be due to differing beam dilution effects.
Instead it is likely to be due to differences in either molecular abundance -- e.g., as a result of a distance dependent X-ray ionisation rate -- or differences in excitation efficiencies -- e.g., due to higher nuclear gas densities favoring HCN. (The critical H$_2$ density of $\sim  7\times 10^{3}$~cm$^{-3}$
for collisional deexcitation of CO (2-1) 
is significantly lower that the $\sim 3\times 10^6$~cm$^{-3}$ for HCN (1-0).)
In the next section we use large velocity gradient (LVG) models to address these issues.

\begin{table}
\caption{CO and HCN fluxes in NGC3227}
\label{tab:data}
\begin{tabular}{l c c c}
\hline\hline
%region  & radius & F$_{HCN1-0}$ & F$_{CO2-1}$ & F$_{HCN1-0}$/F$_{CO2-1}$ \\
%        &        & Jy\,km\,s$^{-1}$ & Jy\,km\,s$^{-1}$ & \\
aperture & F$_{HCN1-0}$      & F$_{CO2-1}$\tablefootmark{a} & F$_{HCN1-0}$/F$_{CO2-1}$\\
         & Jy\,km\,s$^{-1}$ & Jy\,km\,s$^{-1}$ & \\
\hline
1\arcsec & $0.48\pm0.06$ &  $\phantom{0}4.4\pm\phantom{0}1.3$ & $0.114^{+0.051}_{-0.035}$ \\
2\arcsec & $1.37\pm0.17$ & $19.5\pm\phantom{0}5.9$ & $0.074^{+0.033}_{-0.023}$ \\
4\arcsec & $1.86\pm0.29$ & $57.6\pm17.3$ & $0.033^{+0.016}_{-0.011}$ \\
2--4\arcsec & $0.49\pm0.33$ & $38.5\pm18.3$ & $0.013^{+0.024}_{-0.008}$ \\
\hline
\end{tabular}

\tablefoottext{a}{Scaled from the 75\,mJy given by \cite{sch00} in a 8.4\arcsec$\times$8.4\arcsec\ aperture and with an uncertainty of 30\%.}

\end{table}

%______________________________________________________________

\section{Physical Properties of the Molecular Gas}
\label{sec:gasprop}

\subsection{LVG Calculations}
\label{sec:lvg}

\begin{figure*}
\centering
\includegraphics[angle=0,width=\textwidth]{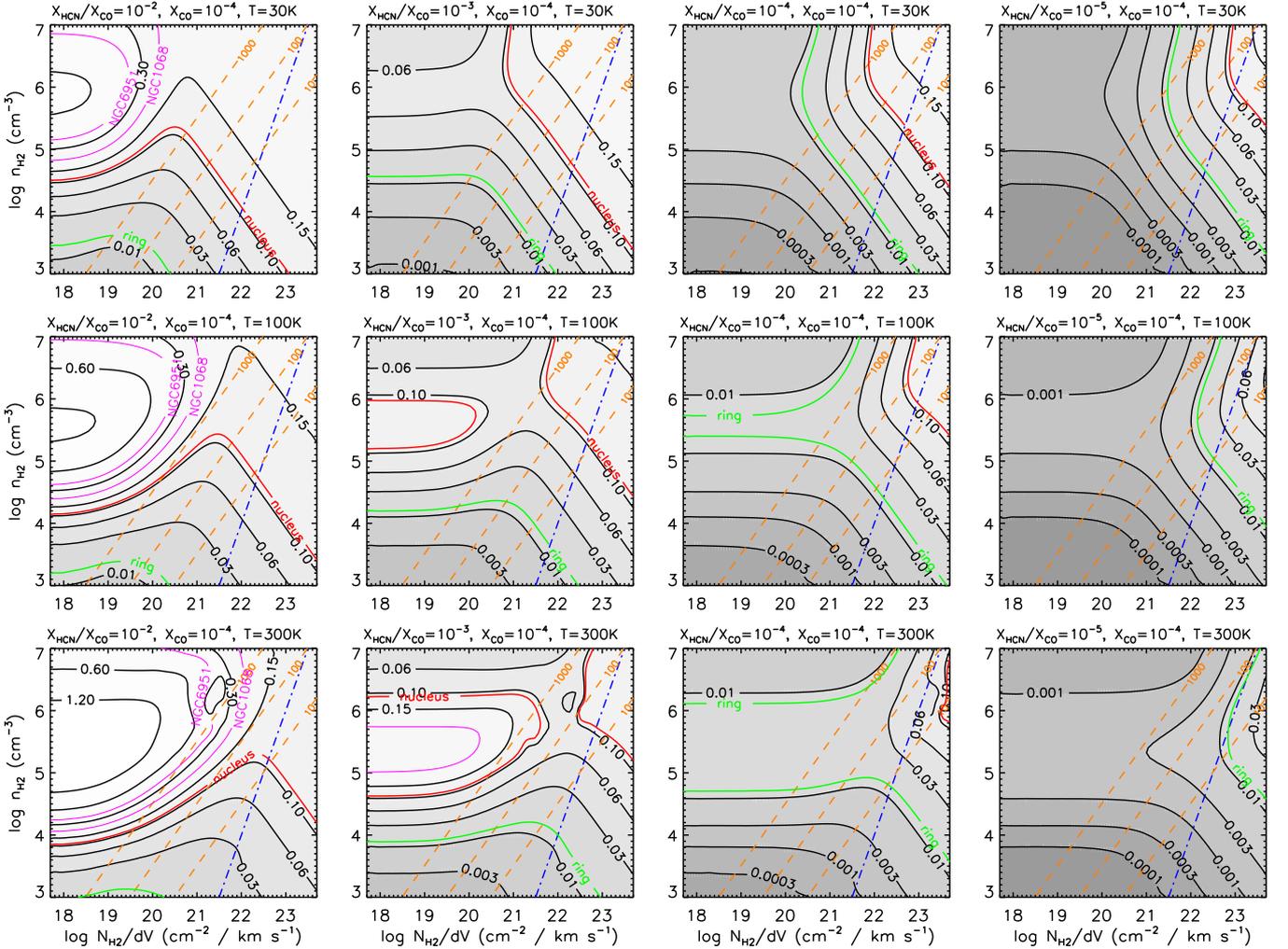}
\caption{\label{fig:lvg}
LVG calculations for a 4-dimensional parameter space: the [HCN]/[CO] abundance ratio and gas kinetic temperature are given at the top of each panel, while the axes of each panel are the gas volume density and ratio between the column density and linewidth (or equivalently between the volume density and velocity gradient). The contours show the expected line ratio based on the emitted fluxes in Jy\,km\,s$^{-1}$.
The locus of representing possible parameters for the nucleus and ring of NGC\,3227 are drawn in red and green respectively.
The dashed orange lines are curves of equal velocity gradient (in km\,s$^{-1}$\,pc$^{-1}$), indicating which regions of the parameter space are physically plausible.
In particular, the locus for self-gravitating virialised clouds is represented by the dot-dash blue line.
}
\end{figure*}

We have constructed a new LVG code 
and have used it to compute the HCN (1-0) and CO (2-1) line intensities for a wide range of conditions.
We calculate the line source functions assuming photon escape probabilities from spherical clouds.  We use the recent \cite{yan10} data for 
the excitations and deexcitations of the CO rotational levels that are induced by collisions with H$_2$.  
For HCN we use the \cite{gre74} collisional data, as updated and listed in
the RADEX database \citep{tak07}.  We have cross-checked all of our results with the LVG code RADEX, 
and find excellent agreement: 
in the parameter space we have assessed, the converged line ratios agree to within 1.5\%.
We assume that the HCN and CO molecules are mixed uniformly, and that the corresponding line emissions arise from gas at the same temperature and density.
The luminosity ratios resulting from the model calculations are presented graphically in Fig.~\ref{fig:lvg}, covering the parameter space:
kinetic temperature $30 \leq T\,[K] \leq 300$,
HCN to CO abundance ratio $10^{-5} \leq X_{HCN}/X_{CO} \leq 10^{-2}$ 
(with $X_{CO}=10^{-4}$ is the CO abundance relative to hydrogen),
H$_2$ volume gas density $10^{3} \leq n_{H_2}\,[cm^{-3}] \leq 10^{7}$,
and a ratio of gas-density to velocity-gradient, or equivalently
column density to linewidth, of  
$5\times10^{17} \leq N_{H_2}/dV\, [cm^{-2}\,(km\,s^{-1})^{-1}] \leq 5\times10^{23}$.

In each panel in Fig.~\ref{fig:lvg}, the parameter-space consists of four regimes.
The upper right parts
% of the panels in Fig.~\ref{fig:lvg} 
correspond to local thermal equilibrium (LTE) in the optically thick limit.
In this regime
%arises because under such conditions 
the transition excitation temperatures $T_{ex}$ of the lines approaches the kinetic temperature $T_{kin}$ of the gas.
Since the line flux, in the Rayleigh-Jeans limit, is 
$F \propto \int T_{ex} \nu^2 d\nu$, and an interval $dV$ in velocity space is 
$dV = c \, d\nu / \nu$, the ratio of two line fluxes measured as 
$\int F \, dV$ (e.g. in units of Jy\,km\,s$^{-1}$ as used here) is 
$F_1/F_2 = (\nu_1/\nu_2)^2$
in the optically thick and LTE limit.
This corresponds to 0.15 for the ratio of the HCN(1-0) and CO(2-1) lines at 88.6\,GHz and 230.5\,GHz respectively,
which is close to the observed ratio of 0.1 in the nucleus.  

The upper left regions also correspond to LTE because the gas densities are high, but 
in this regime the line optical depths are low because the velocity gradients (or line widths) are large.
For such conditions, the line intensities are linearly proportional to the molecular abundances. 
This behavior is  reflected in the HCN/CO intensity ratios indicated by the contour values.
On the left sides of the panels, the intensity ratios decrease linearly with
the assumed abundance ratio
$X_{HCN}/X_{CO}$, which ranges from $10^{-2}$ to $10^{-5}$ in Fig.~\ref{fig:lvg}.

The lower parts of each panel correspond to low densities for which the HCN 
is subthermally excited, leading to relatively higher populations in the lowest rotational levels.
Again the optical depth increases from left to right as the column density increases for fixed line width.

Some regions of the parameter space may be less physically plausible
because they assume very large velocity gradients.
Lines of constant velocity gradient are indicated by the dashed orange lines in Fig.~\ref{fig:lvg},
with increasing gradients towards the upper left.
%These represent curves of equal velocity gradient.
%They trace $dV/dr$ from lower left to upper right, and can be drawn on the plots because 
%$N_{H_2}/dV = n_{H_2}/(dV/dr)$.
For self-gravitating virialised clouds $\Delta V/R \sim n^{1/2}$ where $\Delta V$ is the velocity dispersion, $R$ is the cloud radius,
and $n$ is the gas density. Treating $\Delta V/R$ as a velocity gradient \citep[e.g.][]{gol01} gives 
$dV/dr \sim {\rm 3.1\,km\,s^{-1}\,pc^{-1}} \sqrt{n_{H_2}/10^4\,{\rm cm^{-3}}}$, or
typically a few~km\,s$^{-1}$\,pc$^{-1}$, or a few tens in cases of extreme density.
In  Fig.~\ref{fig:lvg}, the virial relation $n_{H_2} \propto (N_{H_2}/dV)^2$ is represented by the dot-dash blue line in each panel.
Clouds that are unbound or at least partially pressure confined could be to the left of this line.

\subsection{Analysis for NGC\,3227}
\label{sec:nuc}

Curves representing the possible parameter ranges for the measured flux ratios F$_{HCN1-0}$/F$_{CO2-1}$ for NGC\,3227 have been overdrawn for the nucleus (red lines) and circumnuclear ring (green lines), as defined in Sec.~\ref{sec:data}.
If we assume that the clouds are self-gravitating, we are restricted to the points where the red and green lines intersect the dashed blue line.
These show that if the clouds in the ring have a similar density to the nucleus, then the HCN abundance must be about 2 orders of magnitude lower.
Alternatively, the density may differ by up to 2 orders of magnitude if the abundances are similar. 
We discuss this further in Sec.~\ref{sec:xray}

In the nucleus, the flux ratio F$_{HCN1-0}$/F$_{CO2-1}$=0.11 is remarkably close to the LTE optically thick limit, and its locus includes a contour around this region.
However, depending on the HCN abundance, the gas density at which this occurs can vary from $n_{H_2} \sim 10^4$\,cm$^{-3}$ at the highest abundance to 
$\sim 3\times10^5$\,cm$^{-3}$ at the lowest abundance we have considered, 
with column density 
$N_{H_2}/dV \gtrsim 10^{22}$\,cm$^{-2}$/(km\,s$^{-1}$).
By considering the mean volume density of the HCN emitting region, and putting a limit on a realistic filling factor, we can restrict this range further.

The first step is to estimate the volume of the emitting region. This can be done because the HCN emission is marginally resolved.
A detailed estimate of the intrinsic size -- via dynamical modelling, and taking into account emission from the ring -- is given in \cite{san11}.
These authors show that the diameter is 0.54\arcsec, corresponding to 45\,pc, and the scale height is 6\,pc.
Taking this as an indication of the size along the line of sight sets the volume of the emitting region.

The second step is to estimate the mass.
We have done this in several ways since they are all uncertain.
\begin{enumerate}
\item
The LVG calculation yields a mass directly under the assumption that the ratio of the observed linewidth to the linewidth of an individual cloud is tracing the number of clouds, such that 
$N_{tot} = N_{cloud} (\delta\nu_{obs} / \delta\nu_{cloud})$.
For conditions corresponding to 
$X_{HCN}/X_{CO} = 10^{-2}$, 
T$=300$\,K, 
$n_{H_2} = 10^{5.5}$\,cm$^{-3}$ and 
$N_{H_2}/dV = 10^{22}$\,cm$^{-2}$\,(km\,s$^{-1})^{-1}$ (see Sec.~\ref{sec:others}), the observed HCN(1-0) flux leads to a mass of
$1.8\times10^6$\,M$_\odot$.
\item
The standard method to estimate the mass is from the CO luminosity. 
We have done this directly from the CO(2-1) line flux in Table~\ref{tab:data} using a conversion factor $\alpha = 4.3$ which includes a correction for helium \citep{tac08}.
This yields $3.3\times10^6$\,M$_\odot$.
\item
A similar conversion for the HCN line that has been calibrated by \cite{kri08} for AGN is 
$M_{H_2}$/L$_{HCN} \sim 10\,M_\odot$\,(K\,km\,s$^{-1}$\,pc$^2$)$^{-1}$.
This yields a mass of $6\times10^6$\,M$_\odot$.
\item
As a final check, we use the dynamical mass derived from the HCN kinematics.
By fitting models to account for the beam smearing, \cite{san11} find $M_{dyn} = 5.6\times10^7$\,M$_\odot$.
For a nominal 10\% gas fraction expected in local disks and starbursts \citep{hic09}, this would yield a gas mass of $6\times10^6$\,M$_\odot$.
\end{enumerate}

These estimates are all the same order of magnitude, and suggest that the gas mass in the central arcsec is of order $4\times10^6$\,M$_\odot$.
Hence we can estimate the mean density to be 
$\langle n_{H_2}\rangle \gtrsim 6 \times10^3$\,cm$^{-3}$.
Comparing this to the cloud densities above yields volume filling factors in the range 1--0.01.
In this range, a lower filling factor is more physically plausible, 
which would tend to favour the solutions with higher cloud densities.
Fig.~\ref{fig:lvg} shows these have either higher temperature or less extreme HCN abundance.

\subsection{Comparison to NGC\,6951 and NGC\,1068}
\label{sec:others}

NGC\,1068 and NGC\,6951 are two other galaxies for which the HCN(1-0)/CO(2-1) ratio has been measured on comparable $\sim100$\,pc scales.
We use flux densities reported by \cite{kri07} for the nuclear region (denoted `C' in their Table~1) of NGC\,6951; and also the values for the circumnuclear disk of NGC\,1068, as the sum of the red and blue channels reported in Table~3 of \cite{use04}.
These yield line ratios (for line fluxes in Jy\,km\,s$^{-1}$) of $0.37\pm0.05$ and $0.214\pm0.002$ respectively, and are denoted by the solid magenta lines on Fig.~\ref{fig:lvg}.
These lines appear almost exclusively in the panels corresponding to the highest HCN abundance we have considered, $X_{HCN}/X_{CO}=10^{-2}$.

In contrast to NGC\,3227, in which the line emission appears to be optically thick, the loci of the magenta lines for NGC\,1068 and NGC\,6951 are toward the optically thin (left) side of the panels.
Despite this, it is notable that there are regions of the parameter space where the contours corresponding to all 3 objects lie close together, running from lower left to upper right.
The region extends from $n_{H_2}=10^{4}$\,cm$^{-3}$ and $N_{H_2}/dV=10^{19}$\,cm$^{-2}$/(km\,s$^{-1}$)
to $n_{H_2}=10^{6}$\,cm$^{-3}$.
%If we lift the explicit assumption made in Sec.~\ref{sec:nuc} that the clouds are virialised, we can consider the regime where the intensity ratio in NGC\,3227 is closest to that measured in NGC\,1068 and NGC\,6951.
It is precisely because one can attribute the observed line ratios -- with different optical depths for the 3 galaxies -- to similar physical properties of the gas in all these 3 objects that this locus is appealing.

\begin{figure}
\centering
\includegraphics[angle=0,width=9.0cm]{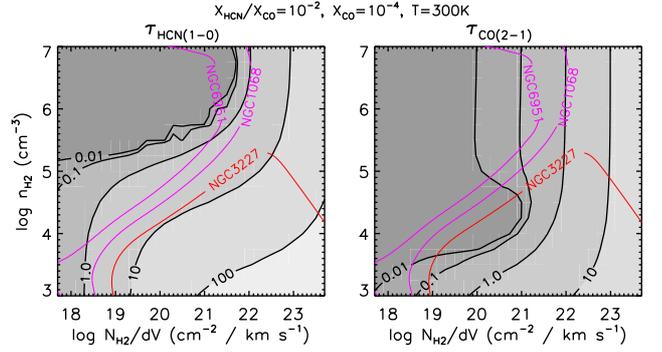}
\caption{\label{fig:tau}
Optical depth $\tau$ of the HCN(1-0) line (left) and CO(2-1) line (right) for gas properties corresponding to the bottom left panel in Fig.~\ref{fig:lvg}.
Darker regions correspond to lower optical depth.
Overplotted are the contours for the 3 galaxies corresponding to their HCN(1-0)/CO(2-1) ratio.}
\end{figure}

Why this occurs can be seen in Fig.~\ref{fig:tau} which shows the optical depths $\tau$ for the HCN(1-0) and CO(2-1) transitions.
The gas properties of both these panels correspond to the bottom left panel in Fig.~\ref{fig:lvg} (300\,K and $X_{HCN}/X_{CO}=10^{-2}$), and cover the same range of density and velocity gradient.
These plots show clearly the characterisation of the different regions:
in the lower half the HCN(1-0) line is optically thick because the density is low enough that it is sub-thermal; above the critical density, the line is in LTE and thus optically thin at low columns and optically thick at high columns.
The locus where all the contours for the 3 galaxies are close together and parallel follows approximately the boundary where the HCN(1-0) line becomes optically thick.
Here, a small change in physical conditions (column or density) can result in the HCN(1-0) emission switching from optically thin to optically thick.

This regime is, however, also associated with very large velocity gradients.
It is $dV/dr\sim10^4$\,km\,s$^{-1}$\,pc$^{-1}$ at $T=30\,K$, but reduces as the temperature increases.
Velocity gradients were not discussed explicitly by \cite{ste94} or \cite{use04} in their $T=50$\,K LVG calculations for NGC\,1068.
But their analyses also associate the observed properties with similarly extreme velocity gradients.
Indeed, one of the main conclusions of \cite{ste94} was that 
$X_{HCN}/X_{CO} \gtrsim 10^{-2}$ in NGC1068.
For the temperature they considered, this would lead to $dV/dr\sim10^4$\,km\,s$^{-1}$\,pc$^{-1}$ (matching the top left panel of Fig.~\ref{fig:lvg} here).
However, our LVG calculations shows that $dV/dr$ is reduced as both the temperature and density increase.
When considering all 3 galaxies together, the smallest -- and therefore arguably the most physically plausible -- value in the parameter space we have covered is 
$dV/dr\sim100$\,km\,s$^{-1}$\,pc$^{-1}$ at 
$T=300$\,K and $n_{H_2}\sim10^{5.5}$\,cm$^{-3}$.
This location is not far from the boundary of the optically thick LTE regime discussed previously, but due to the high velocity gradient represents clouds that are either pressure confined or unbound.

Interestingly, there is evidence in NGC\,1068 from recent Herschel observations with PACS of high rotational CO transitions, for a significant mass of molecular gas in the central $\sim100$\,pc at temperatures of 100\,K and 400\,K and densities of $\sim10^{6.5}$\,cm$^{-3}$ \citep{hai11}.
Similarly, in an analysis of various HCN, HCO$+$ and CO isotope transitions in the central 100\,pc of NGC\,1068, \cite{kri11} also argued in favour of warm (T$\gtrsim200$\,K) gas.
However, they also concluded that the density is of order $n_{H_2}\sim10^{4}$\,cm$^{-3}$.
Our LVG calculations indicate that such densities are associated with very high velocity gradients for the observed line ratio, which we consider physically unlikely.
This, combined with a comparison of the cloud density to mean density estimated in Sec.~\ref{sec:nuc} has led us to favour the higher density solution with more moderate velocity gradient.

\subsection{X-ray Ionisation rate in NGC\,3227}
\label{sec:xray}

%Although our solution cannot currently be fully constrained, the observations and 
Our LVG calculations presented above suggest that as one alternative the high HCN(1-0)/CO(2-1) ratio in the central 
$\sim$100\,pc of NGC\,3227 could be due to an exceptionally high HCN abundance,
an interpretation supported by the even higher HCN/CO intensity ratios observed in the nuclei of
NGC\,1068 and NGC\,6951.
% and NGC\,3227 requires an exceptionally high HCN abundance.
It is possible that the high HCN abundances are associated with elevated X-ray
ionisation and/or heating rates near the AGN. 
Theoretical investigations of X-ray (or cosmic-ray) driven chemistry show that the
(steady-state) molecular abundances depend primarily on the ratio of the
cloud density $n$ to X-ray ionisation rate
$\zeta$ \citep{kro83,lep96,mal96,mei05,bog05,mei07}.
Here we adopt the notation of \cite{bog05},
normalising $\zeta$ to $10^{-17}$\,s$^{-1}$ to give the parameter
$n/\zeta_{-17}$. The density $n = n_H + n_{H_2}$ refers to the total
atomic plus molecular hydrogen density.
In the following analysis, we first calculate the ratio $n/\zeta_{-17}$ in the nucleus of NGC\,3227 and then compare it to ratios predicted by models of X-ray irradiated gas.

\subsubsection{$n/\zeta_{-17}$ in NGC\,3227}

To estimate the X-ray ionisation rate in the central $\sim$100\,pc of NGC 3227
%in this section we explore the specific case of NGC\,3227, to assess whether x-ray ionisation from its central AGN can account for the high HCN abundance.
%To calculate $\zeta_{-17}$ for NGC\,3227, 
we need to know the intrinsic spectral energy distribution from the AGN.
We adopt
the SXPL model of \cite{mar09} in which both the hard and soft
components of the X-ray flux are modelled with power laws:
\[
N_{\rm ph} = 0.0040(E/keV)^{-3.35} + 0.0067(E/keV)^{-1.57}
\]
where $E$ is the photon energy in keV and $N_{\rm ph}$ is the photon
flux in units of ph\,keV$^{-1}$\,cm$^{-2}$\,s$^{-1}$.
%For a distance of 17\,Mpc this yields a luminosity in the 1--100\,keV
%range of $5\times10^{42}$\,erg\,s\,$^{-1}$.
Making the usual assumption that the primary ionisation rate of
hydrogen is negligible compared to the secondary ionisation rate, we
then calculate the resulting ionisation rate $\zeta$ using equation
A4 of \cite{mal96}:
\[
\zeta = N_{\rm sec} \int^{E_{\rm max}}_{E_{\rm min}} 
                    \sigma_{\rm pa}(E) \, F(E) \, dE.
\]
Here $N_{\rm sec}=28$ 
(as given by \citealt{mal96}) is the number of
secondary ionisations per keV of primary photoelectron energy 
assuming a mean-energy per ion-pair of 37.1 eV for 
energy deposition in a 
molecular hydrogen gas (\citealt{dal99});
$\sigma_{\rm pa}(E)$ is the absorption cross section per
H nucleus, for which we adopt the broken power-law fit in equation A5
of \cite{mal96};
and $F(E)$ is the incident flux in units of ph\,keV$^{-1}$.
The photoionisation is dominated by
photons with energies for which $\tau\sim 1$.
We therefore take
the limits of the integral to be $E_{\rm max}=100$\,keV and
$E_{\rm min}$ as the energy at which the optical depth due to
photoelectric absorption is $\tau(E)=1$, ignoring attenuation above
this limit.
The energy at which $\tau(E)=1$ is interpolated from Table~9.3 of
\cite{sew00} for the \cite{mor83} model, for a given column density.
And the column density is assumed to be proportional to distance from
the AGN up to a maximum of $3\times10^{23}$\,cm$^{-2}$ (\citealt{hic09})
at 30\,pc.

We have evaluated the integral at two distances: 18\,pc, an area
weighted mean distance from the AGN to the HCN emitting gas that
corresponds to the nuclear region;
and 140\,pc corresponding to the distance of the circumnuclear ring.
We find 
$\zeta_{\rm 18pc} = 3.6\times10^{-13}$\,s$^{-1}$ and
$\zeta_{\rm 140pc} = 3.5\times10^{-15}$\,s$^{-1}$, about a factor 
100 less primarily due to the distance related geometrical
dilution of the incident X-ray flux. The inferred ionisation rates
are much larger than the typical ionisation rates in 
Galactic clouds.

Adopting a characteristic density
$n_{H_2}\sim10^{5.5}$\,cm$^{-3}$ from our LVG analysis in Sec.~\ref{sec:others}.
yields 
$n/\zeta_{-17} \sim 10$ in the nuclear region of NGC\,3227.
We note that even if the gas density were an order of magnitude higher (leading to proportionally higher $n/\zeta_{-17}$), 
the physical conditions will be well within the high-ionisation phase for the molecular chemistry.

\subsubsection{Models of abundance ratio as a function of $n/\zeta_{-17}$}

\begin{figure}
\centering
\includegraphics[angle=0,width=8.8cm]{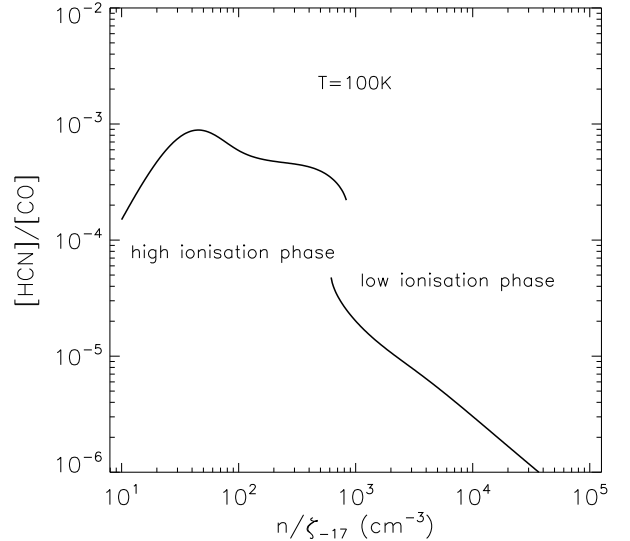}
\caption{\label{fig:ratio}
Abundance ratio [HCN]/[CO] as a function of $n/\zeta_{-17}$, the ratio of the local gas volume density to ionisation rate,
for 100\,K gas, showing the transition from the high- to low-ionisation phase.
}
\end{figure}

The model computations of \cite{bog05,bog06} show that gas can exist in a high (low) ionisation
phase for small (large) values of $n/\zeta_{-17}$, with a density ratio of $n/\zeta_{-17} \sim 10^3$ marking the cross-over between the two regimes.
The HCN/CO abundance ratio can become large  $\gtrsim 10^{-3}$ in the high-ionisation phase even at low gas temperatures, due to the large densities of atomic and ionic carbon. 
%In the high-ionisation phase the HCN/CO abundance can become large $\gtrsim 10^{-3}$ even at  low gas temperatures. 
Here we have re-run these calculations using the same elemental gas-phase abundances, and with a slightly updated reaction set.
Fig.~\ref{fig:ratio} shows the resulting abundance ratios as a function of $n/\zeta_{-17}$ 
for 100\,K gas.
At very low $n/\zeta_{-17} \lesssim 100$ corresponding to that in the nuclear region of NGC\,3227 
the HCN/CO abundance ratio approaches $\sim 10^{-3}$.  For comparison, \cite{lep96} find a peak value of 
$5\times 10^{-4}$ in their calculation.
These models track the abundances of numerous molecules across many orders of magnitude and show that at low $n/\zeta_{-17}$, the HCN/CO abundance ratio is raised several orders of magnitude above that typically expected, to a level at which it approaches -- and, given the uncertainties in such models, is commensurate with -- that implied by the LVG models.

\cite{har10} have shown that at elevated temperatures $\gtrsim$~300~K, rapid hydrogenation of CN to HCN can increase the HCN/CO abundances further still,
and they comment that such warm gas may be present in the X-ray heated gas near AGN.  
We caution however, that their higher temperature
models for which HCN/CO is largest, correspond to $n/\zeta_{-17} = 10^{4.5}$ which is significantly larger than the value we are
invoking as characteristic for the nucleus of NGC\,3227.

The models show that
a combination of X-ray ionisation and heating do yield high HCN abundances, although
not yet quite as high as HCN/CO$\sim10^{-2}$ as
implied by the LVG analysis in Sec.~\ref{sec:others} for the nucleus.
Further chemical modeling is required, but a high HCN/CO intensity ratio
in the nucleus due to elevated X-ray ionisation rates appears plausible.
In this picture, the lower HCN/CO intensity ratio in the ring may simply reflect the lower ionisation rate there,
and not just a lower gas density in this circumgalactic environment.

%______________________________________________________________

\section{Summary and Conclusions}

We present an LVG analysis of high-resolution
observations of CO (2-1) and HCN (1-0) line emissions in the central regions of the Seyfert galaxy NGC 3227. 
%We use large velocity gradient (LVG) calculations to assess the range of physical conditions that are consistent with the observed HCN(1-0)/CO(2-1) flux ratio in the central regions of NGC\,3227. 
We find that 
\begin{itemize}
\item
The HCN(1-0)/CO(2-1) ratio in NGC\,3227 is an order of magnitude higher in the central 80\,pc than in the circumnuclear ring at a radius of 140\,pc.
NGC\,6951 and NGC\,1068 have similarly high published ratios in their central $\sim100$\,pc.
\item
The nuclear HCN/CO intensity ratio in NGC 3227 may reflect optically thick line emission in dense gas with only a weak
constraint on the HCN/CO abundance ratio. However, our
LVG calculations also indicate that the high nuclear ratios in all three of these galaxies are more consistent with
a single set of physical properties corresponding to warm $\sim300$\,K, dense $10^{5.5}$\,cm$^{-3}$ gas, 
in which the emission lines are optically thinner, but
in which the HCN/CO abundance ratio is very large $\sim 10^{-2}$.
For these conditions the velocity gradients are $dV/dr \sim 100$\,km\,s$^{-1}$, but would increase significantly at lower temperatures or densities.
Most likely the clouds are gravitationally unbound.
\item
The X-ray ionisation rate at radii less than $\sim 20$~pc may exceed 10$^{-13}$~s$^{-1}$,
and could plausibly lead to high HCN abundances in 
%warm 
molecular gas in the high-ionisation phase where the ratio of the gas density 
to the X-ray ionisation rate is small.
%Comparison of the x-ray ionisation rate and gas density to models of molecular abundances in x-ray irradiated gas indicates that the high HCN line intensity in the central 100\,pc could be due to the AGN.
\end{itemize}

\begin{acknowledgements}
The authors thank the IRAM staff, in particular Jeremie Boissier and Sascha Trippe, for their invaluable help in obtaining and reducing the data presented in this paper. We thank J.~Graci\'a-Carpio for many useful and interesting discussions. We thank the DFG
for support via German- Israeli Project Cooperation grant STE1869/1-1.GE625/15-1.

\end{acknowledgements}


\begin{thebibliography}{}

\bibitem[Boger \& Sternberg(2005)]{bog05}
Boger G., Sternberg A., 2005, 
ApJ, 632, 302

\bibitem[Boger \& Sternberg(2006)]{bog06}
Boger G., Sternberg A., 2006, 
ApJ, 645, 314

\bibitem[Dalgarno et al.(1999)]{dal99}
Dalgarno, A., Yan, M., Liu, W., 1999
ApJS, 125, 237

\bibitem[Davies et al.(2006)]{dav06}
Davies R., et al., 2006,
ApJ, 646, 754

\bibitem[Davies et al.(2007)]{dav07}
Davies R., M\"uller S\'anchez F., Genzel R., Tacconi L., Hicks E.,
Friedrich S., Sternberg A., 2007,
ApJ, 671, 1388

\bibitem[Goldsmith(2001)]{gol01}
Goldsmith P., 2001,
ApJ, 557, 736

\bibitem[Graci\'a-Carpio et al.(2006)]{gra06}
Graci\'a Carpio J., Garc\'ia Burillo S., Planesas P., Colina L., 2006,
ApJ, 640, L135

\bibitem[Graci\'a-Carpio et al.(2008)]{gra08}
Graci\'a Carpio J., Garc\'ia Burillo S., Planesas P., Fuente A., Usero A., 2008
A\&A, 479, 703

\bibitem[Green \& Thaddeus(1974)]{gre74}
Green, S., Thaddeus, P., 1974
ApJ, 191, 653

\bibitem[Hailey-Dunsheath et al.(2011)]{hai11}
Hailey-Dunsheath S., et al., 2011,
in prep.

\bibitem[Harada et al.(2010)]{har10}
Harada N., Herbst E., Wakelam V., 2010
ApJ, 721, 1570

\bibitem[Hicks et al.(2009)]{hic09}
Hicks E., Davies R., Malkan M., Genzel R., Tacconi L., M\"uller
S\'anchez F., Sternberg A., 2009,
ApJ, 696, 448

\bibitem[Kohno et al.(2001)]{koh01}
Kohno K., Matsushita S., Vila-Vilar\'o B., Okumura S.,
Shibatsuka T., Okiura M., Ishizuki S., Kawabe R., 2001,
in {\em The Central Kiloparsec of Starbursts and AGN},
eds Knapen J., Beckman J., Shlosman I., Mahoney T., 
ASP Conf. Proc., 249, p.672

\bibitem[Kohno(2005)]{koh05}
Kohno K., 2005,
in {\em The Evolution of Starbursts},
eds H\"uttermeister S., Manthey E., Bomans D., Weis K.,
AIP Conf. Proc., vol. 783, p. 203--208

\bibitem[Kohno et al.(2008)]{koh08}
Kohno K., et al., 2008, 
in {\em Far-Infrared Workshop},
eds Kramer C., Aalto S., Simon R.,
EAS Publication Series, vol. 31, p. 65--71

\bibitem[Krips et al.(2007)]{kri07}
Krips M., et al., 2007,
A\&A, 468, L63

\bibitem[Krips et al.(2008)]{kri08}
Krips M., Neri R., Garci\'a Burillo S., Mart\'in S., Combes F.,
Graci\'a Caprio J., Eckart A., 2008,
ApJ, 677, 262

\bibitem[Krips et al.(2011)]{kri11}
Krips M., et al., 2011
ApJ, 736, 37

\bibitem[Krolik \& Kallman(1983)]{kro83}
Krolik J., Kallman T., 1983,
ApJ, 267, 610

\bibitem[Lepp \& Dalgarno(1996)]{lep96}
Lepp S., Dalgarno A., 1996,
A\&A, 306, L21

\bibitem[Maloney et al.(1996)]{mal96}
Maloney P., Hollenbach D., Tielens A., 1996,
ApJ, 466, 561

\bibitem[Markowitz et al.(2009)]{mar09}
Markowitz A., Reeves J., George I., Braito V., Smith R., Vaughan S.,
Ar\'evalo P., Tombesi F., 2009,
ApJ, 691, 922

\bibitem[Meijerink \& Spaans(2005)]{mei05}
Meijerink R., Spaans M., 2005,
A\&A, 436, 397

\bibitem[Meijerink et al(2007)]{mei07}
Meijerink R., Spaans M., Israel F., 2007,
A\&A, 461, 793

\bibitem[Morrison \& McCammon(1983)]{mor83}
Morrison R., McCammon D., 1983,
ApJ, 270, 119

\bibitem[Sani et al.(2011)]{san11}
Sani E., et al., MNRAS submitted

\bibitem[Schinnerer et al.(2000)]{sch00}
Schinnerer E., Eckart A., Tacconi L., 2000,
ApJ, 533, 826

\bibitem[Seward(2000)]{sew00}
Seward F., in {\em Allen's Astrohpysical Quantities, 4th edition}, 
ed. Cox A., 
p. 183

\bibitem[Sternberg et al.(1994)]{ste94}
Sternberg A., Genzel R., Tacconi L., 1994
ApJ, 436, L131

\bibitem[Tacconi et al.(2008)]{tac08}
Tacconi L., et al., 2008,
ApJ, 680, 246

\bibitem[Usero et al.(2004)]{use04}
Usero A., Garci\'a-Burillo S., Fuente A., Martin-Pintado J., Rodr\'iguez-Fern\'andez N., 2004,
A\&A, 419, 897

%\bibitem[Vall\'ee(1995)]{val95}
%Vall\'ee J., 1995,
%Ap\&SS, 234, 1

\bibitem[Van der Tak et al.(2007)]{tak07}
Van der Tak F., Black J., Sch\"oier F., Jansen D., van Dishoeck E., 2007,
A\&A, 468, 627

%\bibitem[Vollmer et al.(2008)]{vol08}
%Vollmer B., Beckert T., Davies R., 2008,
%A\&A, 491, 441

\bibitem[Yang et al.(2010)]{yan10}
Yang B., Stancil P.C., Balakrishnan N., Forrey R.C., 2010,
ApJ, 718, 1062

\end{thebibliography}
\end{document}